\documentclass[11]{article}

\usepackage{latexsym}
\usepackage{amsfonts}
\usepackage{graphicx}
\usepackage{amsmath}
\usepackage{amssymb}
\usepackage[latin1]{inputenc}

\title{\textbf{Bicomplex Quantum Mechanics:\\ II. The Hilbert Space}}
\author{D. Rochon\thanks{E-mail: \texttt{Dominic.Rochon@UQTR.CA}} \and
S. Tremblay\thanks{E-mail: \texttt{Sebastien.Tremblay@UQTR.CA}}}
\date{D\'epartement de math\'ematiques et
d'informatique \\ Universit\'e du Qu\'ebec \`a Trois-Rivi\`eres \\
C.P. 500 Trois-Rivi\`eres, Qu\'ebec \\ Canada, G9A 5H7}

\makeatletter \@addtoreset{equation}{section}

\makeatletter

\def\be   {\begin{equation}}   \def\ee   {\end{equation}}
\def\ba   {\begin{array}}      \def\ea   {\end{array}}
\def\bea  {\begin{eqnarray}}   \def\eea  {\end{eqnarray}}
\def\bean {\begin{eqnarray*}}  \def\eean {\end{eqnarray*}}

\newtheorem{theorem} {Theorem}
\newtheorem{lemma}{Lemma}
\newtheorem{definition} {Definition}

\newtheorem{corollary} {Corollary}

\newcommand{\pre}{\mathrm{Re}}
\newcommand{\pim}{\mathrm{Im}}

\newcommand{\bi} {\ensuremath{{\bf i}}}
\newcommand{\bo} {\ensuremath{{\bf i_1}}}
\newcommand{\bos}{\ensuremath{{\bf i_1^{\text 2}}}}
\newcommand{\bt} {\ensuremath{{\bf i_2}}}
\newcommand{\bts}{\ensuremath{{\bf i_2^{\text 2}}}}

\newcommand {\bj}{\ensuremath{{\bf j}}}
\newcommand {\bjs}{\ensuremath{{\bf j^{\text 2}}}}
\newcommand{\eo} {\ensuremath{{\bf e_1}}}
\newcommand{\et} {\ensuremath{{\bf e_2}}}

\newcommand{\bra}[1]{\langle #1|}
\newcommand{\ket}[1]{|#1\rangle}
\newcommand{\braket}[2]{\langle #1|#2\rangle}

\newcommand{\mC}{\ensuremath{\mathbb{C}}}
\newcommand{\mD}{\ensuremath{\mathbb{D}}}

\newcommand{\mR}{\ensuremath{\mathbb{R}}}
\newcommand{\mT}{\ensuremath{\mathbb{T}}}

\begin{document}
\maketitle
\begin{abstract}
\large Using the bicomplex numbers $\mathbb{T}\cong {\rm Cl}_{\Bbb{C}}(1,0) \cong {\rm Cl}_{\Bbb{C}}(0,1)$
which is a commutative ring with zero divisors defined by
$\mathbb{T}=\{w_0+w_1 {\bf i_1}+w_2{\bf i_2}+w_3 {\bf j}\ |\ w_0,w_1,w_2,w_3 \in
\mathbb{R}\}$ where ${\bf i_1^{\text 2}}=-1,\ {\bf i_2^{\text 2}}=-1,\ {\bf j^{\text 2}}=1 \mbox{ and }\ {\bf i_1}{\bf i_2}={\bf j}={\bf i_2}{\bf i_1}$, we construct hyperbolic and bicomplex Hilbert
spaces. Linear functionals and dual spaces are considered on these
spaces and properties of linear operators are obtained; in
particular it is established that the eigenvalues of a
bicomplex self-adjoint operator are in the set of
hyperbolic numbers.
\end{abstract}

\vspace{4cm}
\noindent \textbf{Keywords: }Bicomplex Numbers, Hyperbolic Numbers, Complex Clifford Algebras, Generalized Quantum Mechanics, Hilbert Spaces, Free Modules,
 Linear Functionals, Self-Adjoint Operators.\\

\normalsize
\newpage

\section{Introduction}
Many papers have been written on the extension of the formalism of
quantum mechanics. These generalizations have been done mainly
over quaternions or over the Cayley algebra (octonions), see for
instance \cite{1, 2, 3, 4}. The reason why people have worked
mainly on this algebraic structures to generalize quantum
mechanics comes from the fact that there exist only four normed
division algebra \cite{5}: real ($\mathbb{R}$), complex
($\mathbb{C}$), quaternions ($\mathbb{H}$) and Cayley algebra
($\mathbb{O}$). Cayley algebra has an important blank since
associativity is crucial. Indeed, in \cite{1} it is shown that
quantum mechanics cannot be formulated over the Cayley algebra
since, in a last two instances, associativity is needed for the
existence of Hilbert space. Quantum mechanics over quaternions
seems to work better \cite{1, 2, 3, 16}. However, recently some
interest have been deployed to study quantum mechanics for
associative and commutative algebras beyond the paradigm of
algebras without zero divisors \cite{6, 7, 8}. This leads to a
wide spectrum of possibilities, among which we have the hyperbolic
numbers $\mathbb{D}\cong {\rm Cl}_{\Bbb{R}}(0,1)$ (also called duplex numbers) \cite{9}, the
bicomplex numbers $\mathbb{T}\cong {\rm Cl}_{\Bbb{C}}(1,0) \cong {\rm Cl}_{\Bbb{C}}(0,1)$
\cite{11} and, more generally, the
multicomplex numbers \cite{10, 17}.

In recent years, theory of bicomplex numbers and bicomplex
functions has found many applications, see for instance
\cite{18,19,20,21,22}. Bicomplex numbers is a commutative ring
with unity which contains the field of complex numbers and the
commutative ring of hyperbolic numbers.  Bicomplex (hyperbolic)
numbers are \emph{unique} among the complex (real) Clifford
algebras in that they are commutative but not division algebras.
In fact, bicomplex numbers generalize (complexify) hyperbolic
numbers. Note that Hilbert spaces over hyperbolic numbers that
have been studied in \cite{7, 8} and \cite{12} are different
from the hyperbolic Hilbert space that we consider in this paper.

%By using bicomplex numbers instead of quaternions in this paper we
%recover the commutativy property, but lost invertibility for some
%elements on the null-cone \cite{7}. Such an endeavor is
%particularly timely since the fundamental theory of bicomplex
%analysis has now been considerably developed over the last fifteen
%years \cite{5,6}.

In Section~2 we give an overview of the fundamental theory of
bicomplex analysis necessary for this article. Section~3 is
devoted to free modules over the ring of bicomplex numbers (which is not a $C^*$-algebra).
A fundamental result useful for the rest of the paper is presented:
the unique decomposition of any elements of our free module $M$
into two elements of a standard (complex) vector space in terms of
the idempotent basis. The Section~4 (and 5) introduces the
bicomplex scalar product (the hyperbolic scalar product). In
particular, it is shown that one can constructs a metric space
from $M$ and our bicomplex scalar product. In Section~6, we define
the bicomplex Hilbert space; two examples are given. Section~7
introduces the dual space $M^*$ and re-examines the previous
Sections in terms of the Dirac notation. Finally, Section~8
concerns linear operators or more specifically adjoint and
self-adjoint operators as well as the bicomplex eigenvectors
equation.

\newpage
\section{Preliminaries}
Bicomplex numbers are defined as \cite{11, 10, 13}

\be \mathbb{T}:=\{z_1+z_2\bt\ |\ z_1, z_2 \in \mathbb{C}(\bo) \},
\label{enstetra}
\ee where the imaginary units $\bo, \bt$ and $\bj$
are governed by the rules: $\bos=\bts=-1$, $\bjs=1$ and \be
\ba{rclrcl}
   \bo\bt &=& \bt\bo &=& \bj,  \\
   \bo\bj &=& \bj\bo &=& -\bt, \\
   \bt\bj &=& \bj\bt &=& -\bo. \\
\ea \ee Where we define $\mC(\bi_k):=\{x+y\bi_k\ |\ \bi_k^2= -1$
and $x,y\in \mR \}$ for $k=1,2$. Hence it is easy to see that the
multiplication of two bicomplex numbers is commutative.

It is also convenient to write the set of bicomplex numbers as
\be
   \mathbb{T}:=\{w_0+w_1\bo+w_2\bt+w_3\bj\ |\ w_0,
    w_1,w_2,w_3 \in \mathbb{R}\}.
\ee
%with the product of imaginary units given by:
%and  Hence, the bicomplex numbers are commutative. We define
%the following two subsets
%$\mathbb{C}(\mathbf{i_k}) \subset \mathbb{T}$ for $k=1,2$, by
%$\mathbb{C}(\mathbf{i_k}):=\{x+y\mathbf{i_k}|\mathbf{i^{\text 2}_k}=-1
%\mbox{ and } x,y \in \mathbb{R}\}$.
%It is also convenient to write the set of bicomplex numbers as \be
%\mathbb{T}=\{z_1+z_2\mathbf{i_2}|\ z_1,z_2 \in
%\mathbb{C}(\mathbf{i_1})\}. \ee
In particular, in equation (\ref{enstetra}), if we put $z_1=x$ and
$z_2=y\bo$ with $x,y \in \mathbb{R}$, then we obtain the
subalgebra of hyperbolic numbers: $\mathbb{D}=\{x+y\bj\ |\
\bjs=1,\ x,y\in \mathbb{R}\}$.

Complex conjugation plays an important role both for algebraic and
geometric properties of $\mathbb{C}$, as well as in the standard
quantum mechanics. For bicomplex numbers, there are three possible
conjugations. Let $w\in \mathbb{T}$ and $z_1,z_2 \in
\mathbb{C}(\mathbf{i_1})$ such that $w=z_1+z_2\mathbf{i_2}$. Then we
define the three conjugations as:

\begin{subequations}
\label{eq:dag}
\begin{align}
w^{\dag_{1}}&=(z_1+z_2\bt)^{\dag_{1}}:=\overline z_1+\overline z_2
\bt,
\\
w^{\dag_{2}}&=(z_1+z_2\bt)^{\dag_{2}}:=z_1-z_2 \bt,
\\
w^{\dag_{3}}&=(z_1+z_2\bt)^{\dag_{3}}:=\overline z_1-\overline z_2
\bt,
\end{align}
\end{subequations}
where $\overline z_k$ is the standard complex conjugate of complex
numbers $z_k \in \mathbb{C}(\mathbf{i_1})$. If we say that the
bicomplex number $w=z_1+z_2\bt=w_0+w_1\bo+w_2\bt+w_3\bj$ has the
``signature'' $(++++)$, then the conjugations of type 1,2 or 3 of
$w$ have, respectively, the signatures $(+-+-)$, $(++--)$ and
$(+--+)$. We can verify easily that the composition of the
conjugates gives the four-dimensional abelian Klein group:
\begin{center}
\be
\begin{tabular}{|c||c|c|c|c|}
\hline
$\circ$ & $\dag_{0}$ & $\dag_{1}$  & $\dag_{2}$  & $\dag_{3}$  \\
\hline
\hline
$\dag_{0}$    & $\dag_{0}$ & $\dag_{1}$  & $\dag_{2}$  & $\dag_{3}$  \\
\hline
$\dag_{1}$    & $\dag_{1}$ & $\dag_{0}$ & $\dag_{3}$  & $\dag_{2}$ \\
\hline
$\dag_{2}$    & $\dag_{2}$ & $\dag_{3}$  & $\dag_{0}$ & $\dag_{1}$ \\
\hline
$\dag_{3}$    & $\dag_{3}$ & $\dag_{2}$ & $\dag_{1}$ & $\dag_{0}$ \\
\hline
\end{tabular}
\label{eq:groupedag} \ee
\end{center}
where $w^{\dag_{0}}:=w\mbox{ } \forall w\in \mathbb{T}$.

The three kinds of conjugation all have some of the standard properties of
conjugations, such as:
\begin{eqnarray}
(s+ t)^{\dag_{k}}&=&s^{\dag_{k}}+ t^{\dag_{k}},\\
\left(s^{\dag_{k}}\right)^{\dag_{k}}&=&s, \\
\left(s\cdot t\right)^{\dag_{k}}&=&s^{\dag_{k}}\cdot t^{\dag_{k}},
\end{eqnarray}
for $s,t \in \mathbb{T}$ and $k=0,1,2,3$.\\

We know that the product of a standard complex number with its
conjugate gives the square of the Euclidean metric in
$\mathbb{R}^2$. The analogs of this, for bicomplex numbers, are
the following. Let $z_1,z_2 \in \mathbb{C}(\bt)$ and
$w=z_1+z_2\bt\in \mathbb{T}$, then we have that \cite{11}:
\begin{subequations}
\begin{align}
|w|^{2}_{\bo}&:=w\cdot w^{\dag_{2}}=z^{2}_{1}+z^{2}_{2} \in
\mathbb{C}(\bo),
\\*[2ex] |w|^{2}_{\bt}&:=w\cdot
w^{\dag_{1}}=\left(|z_1|^2-|z_2|^2\right)+2\pre(z_1\overline
z_2)\bt \in \mathbb{C}(\bt), \\*[2ex] |w|^{2}_{\bj}&:=w\cdot
w^{\dag_{3}}=\left(|z_1|^2+|z_2|^2\right)-2\pim(z_1\overline
z_2)\bj \in \mathbb{D},
\end{align}
\end{subequations}
where the subscript of the square modulus refers to the subalgebra
$\mathbb{C}(\bo), \mathbb{C}(\bt)$ or $\mathbb{D}$ of $\mathbb{T}$
in which $w$ is projected.

Note that for $z_1,z_2 \in \mathbb{C}(\bo)$ and $w=z_1+z_2\bt\in
\mathbb{T}$, we can define the usual (Euclidean in $\mR^4$) norm
of $w$ as $|w|=\sqrt{|z_1|^2+|z_2|^2}=\sqrt{\pre(|w|^{2}_{\bj})}$.

It is easy to verify that $w\cdot \displaystyle
\frac{w^{\dag_{2}}}{|w|^{2}_{\bo}}=1$. Hence, the inverse of $w$
is given by \be w^{-1}= \displaystyle
\frac{w^{\dag_{2}}}{|w|^{2}_{\bo}}. \ee From this, we find that
the set $\mathcal{NC}$ of zero divisors of $\mathbb{T}$, called
the {\em null-cone}, is given by $\{z_1+z_2\bt\ |\
z_{1}^{2}+z_{2}^{2}=0\}$, which can be rewritten as \be
\mathcal{NC}=\{z(\bo\pm\bt)|\ z\in \mathbb{C}(\bo)\}. \ee
\smallskip\smallskip\smallskip\smallskip

\noindent Let us now recall the following three \textit{real
moduli} (see \cite{11} and \cite{13}):

\begin{enumerate}
\item[\textbf{1)}] For $s,t\in \mathbb{T}$, we define the first
modulus as $|\cdot|_{\bold{1}}:=\big||\cdot|_{\bo}\big|$. This
modulus has the following properties:
\begin{enumerate}
\item[a)] $|\cdot|_{\bold{1}}: \mathbb{T}\rightarrow \mathbb{R}$;
\item[b)] $|s|_{\bold{1}}\ge 0$ with $|s|_{\bold{1}}=0$ iff $s\in\mathcal{NC}$;
\item[c)] $|s\cdot t|_{\bold{1}}=|s|_{\bold{1}}\cdot |t|_{\bold{1}}$.
\end{enumerate}
From this definition, we can rewrite this real pseudo-modulus in a
much practical way as
$$
   |w|_{\bold{1}}=|z^{2}_{1}+z^{2}_{2}|^{1/2}
$$
or
$$
|w|_{\bold{1}}=\sqrt[4]{ww^{\dag_1}w^{\dag_2}w^{\dag_3}}.
$$

%for $w=z_1+z_2\mathbf{i_2}$ with $z_1,z_2 \in
%\mathbb{C}(\mathbf{i_1})$. Moreover, it is also useful to express
%$|\cdot|_{\mathbf{1}}$, in terms of the three bicomplex conjugates,
%i.e.
%\be |w|_{\mathbf{1}}=\sqrt[4]{ww^{\dag_1}w^{\dag_2}w^{\dag_3}}. \label{eq:wr1.2} \ee

\item[\textbf{2)}] For $s,t\in \mathbb{T}$, we can define formally
the second real modulus as
$|\cdot|_{\bold{2}}:=\big||\cdot|_{\bt}\big|$.  But an easy
computation leads to
% This
%modulus has the same properties as $|\cdot|_{\mathbf{1}}$. Indeed we
%can rewrite $|w|_{\mathbf{2}}$ as
\be
   |w|_{\bold{2}}=|w|_{\bold{1}}=|z^{2}_{1}+z^{2}_{2}|^{1/2},
\ee
% $w=z_1+z_2\mathbf{i_1}$ with $z_1,z_2 \in \mathbb{C}(\mathbf{i_2})$.
% Hence, the first and the second pseudo-modulus are the same.
meaning that there are no reasons to introduce $|\cdot|_{\bold{2}}$.

\item[\textbf{3)}]
%For $s,t\in \mathbb{T}$, we
One more option is to define the third modulus as
$|\cdot|_{\bold{3}}:=\big||\cdot|_{\bj}\big|$. It has the
following properties:
\begin{enumerate}
\item[a)] $|\cdot|_{\bold{3}}: \mathbb{T}\rightarrow \mathbb{R}$;
\item[b)] $|s|_{\bold{3}}\ge 0$ with $|s|_{\bold{3}}=0$ iff $s=0$;
\item[c)]
$|s+t|_{\bold{3}}\leq |s|_{\bold{3}}+|t|_{\bold{3}}$;
\item[d)] $|s\cdot
t|_{\bold{3}}\leq\sqrt{2}|s|_{\bold{3}}\cdot
           |t|_{\bold{3}}$;
\item[e)] $|\lambda \cdot t|_{\bold{3}} = |\lambda| \cdot |t|_{\bold{3}}$, for
$\lambda \in \mathbb{C}(\bo)\mbox{ or }\mathbb{C}(\bt).$
\end{enumerate}
Hence $| \cdot |_{\bold{3}}$ determines a structure of a real normed
algebra on $\mT$.  What is more, one gets directly that

%We note that
%\begin{enumerate}
%\item[(i)]
%$|w|_{\mathbf{j}}=|z_1-z_2\mathbf{i_1}|\mathbf{e_1}+|z_1+z_2\mathbf{i_1}|\mathbf{e_2}%\in\mathbb{D}$
%$\forall w=z_1+z_2\mathbf{i_2}\in\mathbb{T}$, \item[(ii)] $|s\cdot
%t|_{\mathbf{j}}=|s|_{\mathbf{j}}|t|_{\mathbf{j}}\mbox{ }\forall
%s,t\in\mathbb{T}$.
%\end{enumerate}
%\noindent From this definition, we can rewrite the modulus
%$|\cdot|_{\mathbf{3}}$ as
\be
    |w|_{\bold{3}}=\sqrt{|z_1|^2+|z_2|^2},
    \label{eq:wr3.1}
\ee for $w=z_1+z_2\bt$ with $z_1,z_2 \in \mathbb{C}(\bo)$, i.e.,
in fact this is just the Euclidean metric in $\mR^4$ written in a
form compatible with the multiplicative structure of bicomplex
numbers.

Note also that
\begin{enumerate}
  \item[(i)] $|w|_\bj=|z_1-z_2\bo|\eo + |z_1+z_2\bo|\et \in \mD,\qquad
             \forall w=z_1 +z_2\bt \in \mT$,
  \item[(ii)] $|s\cdot t|_\bj=|s|_\bj|t|_\bj \qquad \forall s,t \in \mT$.
\end{enumerate}

%Hence, we see that in fact
%$|\cdot|_{\mathbf{3}}$ is simply the Euclidean metric of
%$\mathbb{R}^{4}$, i.e. \be
%|w|_{\mathbf{3}}=|w|=\sqrt{\pre(|w|^{2}_{\mathbf{j}})}.
%\ee
\end{enumerate}

Finally, let us mention that any bicomplex numbers can be written
using an orthogonal idempotent basis defined by
$$
\eo=\frac{1+\bj}{2}\ \ \ \mbox{and}\ \ \
\et=\frac{1-\bj}{2},
$$
where $\bold{e^{\text{2}}_1}=\bold{e_1}$,
$\bold{e^{\text{2}}_2}=\bold{e_2}$, $\bold{e_1}+\bold{e_2}=1$ and
$\bold{e_1}\bold{e_2}=0=\bold{e_2}\bold{e_1}$. Indeed, it is easy
to show that for any $z_1+z_2\bt\in \mathbb{T}$, $z_1,z_2\in
\mathbb{C}(\bo)$, we have
\begin{equation}
z_1+z_2\bt=(z_1-z_2\bo)\eo+(z_1+z_2\bo)\et.
\label{idempotent}
\end{equation}

\section{$\mathbb{T}$-Module}

The set of bicomplex number is a commutative ring. So, to define a
kind of vector space over $\mathbb{T}$, we have to deal with the
algebraic concept of module. We denote $M$ as a free
$\mathbb{T}$-module with the following finit $\mathbb{T}$-basis
$\Big\{\widehat{m}_l \mid l\in \{1,\ldots,n\}\Big\}$. Hence,
$$
M=\left\{\sum_{l=1}^{n}{x_{l}\widehat{m}_l} \mid x_{l}\in\mathbb{T}\right\}.
$$
Let us now define
\begin{equation}
V:=\left\{\sum_{l=1}^{n}{x_{l}\widehat{m}_l} \mid
x_{l}\in\mC(\bo)\right\}\subset M. \label{V}
\end{equation}
The set $V$ is a free
 $\mC(\bo)$-module which depends on a given $\mathbb{T}$-basis of $M$. In fact, $V$ is a complex vector space of
dimension $n$ with the basis $\Big\{\widehat{m}_l \mid l\in
\{1,\ldots,n\}\Big\}$. For a complete traitement of the Module
Theory, see \cite{14}.

\begin{theorem}
Let $\widehat{X}=\displaystyle
\sum_{l=1}^{n}{x_{l}\widehat{m}_{l}},\mbox{ }
x_{l}\in\mathbb{T},\mbox{ }\forall l\in\{1,\ldots,n\}$. Then,
there exist $\widehat{X}_{\eo},\widehat{X}_{\et}\in V$ such that
$$\widehat{X}=\eo \widehat{X}_{\eo} + \et \widehat{X}_{\et}.$$
\label{theo:Xe1+Xe2}
\end{theorem}
\noindent \emph{Proof.} From equation (\ref{idempotent}), it is
always possible to decompose a bicomplex number in term of the
idempotent basis. So let us write $x_l=x_{1l}\eo+x_{2l}\et$ where
$x_{1l},x_{2l}\in\mC(\bo)$, for all $l\in\{1,\ldots,n\}$. Hence,
\bean
\widehat{X} &=& \sum_{l=1}^{n}{x_{l}\widehat{m}_l} = \sum_{l=1}^{n}{(x_{1l}\eo+x_{2l}\et)\widehat{m}_l}= \eo\sum_{l=1}^{n}{(x_{1l}\widehat{m}_l)} + \et\sum_{l=1}^{n}{(x_{2l}\widehat{m}_l)}\\
%  &=& \sum_{l=1}^{n}{(x_{1l}\eo)\widehat{m}_l+(x_{2l}\et)\widehat{m}_l}\\
%  &=& \sum_{l=1}^{n}{(\eo x_{1l})\widehat{m}_l+(\et x_{2l})\widehat{m}_l}\\
%  &=& \sum_{l=1}^{n}{\eo(x_{1l}\widehat{m}_l)+\et(x_{2l}\widehat{m}_l)}\\
%  &=& \sum_{l=1}^{n}{\eo(x_{1l}\widehat{m}_l)} + \sum_{l=1}^{n}{\et(x_{2l}\widehat{m}_l)}\\
  &=& \eo \widehat{X}_{\eo}+ \et \widehat{X}_{\et}\\
\eean where $\widehat{X}_{\bold{e_k}}:=\displaystyle
\sum_{l=1}^{n}{(x_{kl}\widehat{m}_l)}$ for $k=1,2$.
$\Box$\\

\begin{corollary}
The elements $\widehat{X}_{\eo}$ and $\widehat{X}_{\et}$ are
uniquely determined. In other words, $\eo \widehat{X}_{\eo} + \et
\widehat{X}_{\et}=\eo \widehat{Y}_{\eo} + \et \widehat{Y}_{\et}$
if and only if $\widehat{X}_{\eo}=\widehat{Y}_{\eo}$ and
$\widehat{X}_{\et}=\widehat{Y}_{\et}$. \label{coro:xe1xe2}
\end{corollary}
\emph{Proof}. If $\eo \widehat{X}_{\eo} + \et
\widehat{X}_{\et}=\eo \widehat{Y}_{\eo} + \et \widehat{Y}_{\et}$,
then we have $\eo (\widehat{X}_{\eo}-\widehat{Y}_{\eo}) + \et
(\widehat{X}_{\et}-\widehat{Y}_{\et})=\widehat{0}$. Suppose now
that $\Big\{\widehat{m}_{l} \mid l\in \{1,\ldots,n\}\Big\}$ is a
free basis of $M$, then we have
$\widehat{X}_{\bold{e_k}}=\displaystyle
\sum_{l=1}^{n}{x_{kl}\widehat{m}_{l}}$ and
$\widehat{Y}_{\bold{e_k}}=\displaystyle
\sum_{l=1}^{n}{y_{kl}\widehat{m}_{l}}$ ($k=1,2$),
$x_{kl},y_{kl}\in \mathbb{C}(\bo)$. Therefore, we find
$$
\begin{array}{rcl}
\widehat{0}&=&\eo (\widehat{X}_{\eo}-\widehat{Y}_{\eo}) + \et
(\widehat{X}_{\et}-\widehat{Y}_{\et})
\\*[2ex]
&=&\eo \left(\displaystyle
\sum_{l=1}^{n}{x_{1l}\widehat{m}_{l}}-\displaystyle
\sum_{l=1}^{n}{y_{1l}\widehat{m}_{l}}\right)+\et
\left(\displaystyle
\sum_{l=1}^{n}{x_{2l}\widehat{m}_{l}}-\displaystyle
\sum_{l=1}^{n}{y_{2l}\widehat{m}_{l}}\right)
\\*[2ex]
&=&\displaystyle \sum_{l=1}^{n}(x_l-y_l)\widehat{m}_l,
\end{array}
$$
where $x_l:=\eo x_{1l}+\et x_{2l}\in \mathbb{T}$ and $y_l:=\eo
y_{1l}+\et y_{2l}\in \mathbb{T}$. This implies that $x_l=y_l$ for
all $l\in \{1,\ldots,n\}$; in other words $x_{1l}=y_{2l}$ and
$x_{2l}=y_{2l}$, i.e.
$\widehat{X}_{\bold{e_k}}=\widehat{Y}_{\bold{e_k}}$ for
$k=1,2$.

Conversely, if $\widehat{X}_{\eo}=\widehat{Y}_{\eo}$ and
$\widehat{X}_{\et}=\widehat{Y}_{\et}$ we find trivially the
desired result.~$\Box$

\noindent Whenever $\widehat{X}\in M$, we define the projection
$P_{k}:M\longrightarrow V$ as
\begin{equation}
P_{k}(\widehat{X}):=\widehat{X}_{\bold{e_k}} \label{projection}
\end{equation}
for $k=1,2$. This definition is a generalization of the mutually
complementary projections $\{P_{1},P_{2}\}$ defined in \cite{11}
on $\mathbb{T}$, where $\mathbb{T}$ is considered as the canonical
$\mathbb{T}$-module over the ring of bicomplex numbers. Moreover,
from the Corollary \ref{coro:xe1xe2}, $\widehat{X}_{\eo}$ and
$\widehat{X}_{\et}$ are uniquely determined from a given
$\mathbb{T}$-basis and the projections $P_{1}$ and $P_{2}$
satisfies the following property:
\begin{equation}
P_{k}(w_1 \widehat{X}+w_2
\widehat{Y})=P_{k}(w_1)P_{k}(\widehat{X})+P_{k}(w_2)P_{k}(\widehat{Y})\label{projection_prop}
\end{equation}
$\forall w_1,w_2\in\mathbb{T},\mbox{ }\forall
\widehat{X},\widehat{Y}\in M$ and $k=1,2$.

The vector space $V$ is defined from the free $\mathbb{T}$-module
$M$ with a given $\mathbb{T}$-basis. The next theorem tell us that
$M$ is isomorphic to $V^2=\{(\widehat{X};\widehat{Y})\ |\
\widehat{X},\widehat{Y}\in V\}$, where the addition $+_{V^2}$ and
the multiplication $\cdot_{V^2}$ by a scalar are defined by
$$
\begin{array}{rrcl}
+_{V^2}:& V^2\times V^2 &\rightarrow & V^2
\\
 &\Big((\widehat{X}_1;\widehat{Y}_1),(\widehat{X}_2;\widehat{Y}_2)\Big)&\mapsto &
(\widehat{X}_1;\widehat{Y}_1)+_{V_2}(\widehat{X}_2;\widehat{Y}_2)\\
&&&:=(\widehat{X}_1+\widehat{X}_2;\widehat{Y}_1+\widehat{Y}_2),
\\*[2ex]
 \cdot_{V^2}:& \mathbb{T}\times V^2 &\rightarrow& V^2 \\
  & (\lambda,(\widehat{X};\widehat{Y}))&\mapsto& \lambda\cdot_{V^2}(\widehat{X};\widehat{Y})\\
&&&:=(\lambda_1
  \widehat{X};\lambda_2
  \widehat{Y}),
\end{array}
$$
where $\lambda=\lambda_1 \bold{e_1}+\lambda_2
  \bold{e_2}$. Here the symbol $+$ denotes the addition on $V$ and $\lambda_1
  \widehat{X}$ or $\lambda_2 \widehat{Y}$ denotes the multiplication by a scalar on $V$
(which are also the addition and the multiplication defined on
$M$). Note that we use the notation $(\widehat{X};\widehat{Y})$ to
denote an element of $V^2$, instead of the usual notation
$(\widehat{X},\widehat{Y})$, to avoid confusion with the bicomplex
scalar product defined below.

\begin{theorem}
The set $V^2$ defined with the addition $+_{V_2}$ and the
multiplication by a scalar $\cdot_{V^2}$ over the bicomplex
numbers $\mathbb{T}$ is isomorphic to $M$, i.e.
$$
(V^2,+_{V_2},\cdot_{V^2})\simeq (M,+,\cdot).
\label{isoV2M}
$$
\end{theorem}
\emph{Proof}. It is first easy to show that $V^2$ is a
$\mathbb{T}$-module with $+_{V^2}$ and $\cdot_{V^2}$ defined
above. Now let us consider the function $\Phi:V^2\rightarrow M$
defined by $\Phi\big((\widehat{X};\widehat{Y})\big)=\eo
\widehat{X}+ \et \widehat{Y}$. It is not difficult to show that
$\Phi\big((\widehat{X}_1;\widehat{Y}_1)+_{V^2}
(\widehat{X}_2;\widehat{Y}_2)\big)=\Phi\big((\widehat{X}_1;\widehat{Y}_1)\big)+\Phi\big((\widehat{X}_2;\widehat{Y}_2)\big)$
and that $\Phi(\lambda\cdot_{V^2}\widehat{X})=\lambda
\Phi(\widehat{X})$, i.e. that $\Phi$ is an homomorphism. The
function $\Phi$ is a one-to-one function. Indeed if
$\Phi(\big(\widehat{X}_1;\widehat{Y}_1)\big)=\Phi\big((\widehat{X}_2;\widehat{Y}_2)\big)$,
then $\eo \widehat{X}_1+ \et \widehat{Y}_1=\eo \widehat{X}_2+ \et
\widehat{Y}_2$ which implies that $\widehat{X}_1=\widehat{X}_2$
and $\widehat{Y}_1=\widehat{Y}_2$ from Corollary
\ref{coro:xe1xe2}. Finally, $\Phi$ is an onto function since for
all $\widehat{X}=\eo \widehat{X}_{\eo} + \et \widehat{X}_{\et} \in
M$, we have
$\Phi\big((\widehat{X}_{\eo};\widehat{X}_{\et})\big)=\widehat{X}$.
$\Box$

\begin{theorem}
Let $\Big\{\widehat{v}_{l} \mid l\in \{1,\ldots,n\}\Big\}$ a basis
of the vector space $V$ over $\mC(\bo)$. Then
$\Big\{(\widehat{v}_l;\widehat{v}_l) \mid l\in
\{1,\ldots,n\}\Big\}$ is a basis of the free $\mathbb{T}$-module
$(V^2,+_{V^2},\cdot_{V^2})$ and $\Big\{\widehat{v}_{l} \mid l\in
\{1,\ldots,n\}\Big\}$ is a $\mathbb{T}$-basis of $M$.
\label{V2basis}
\end{theorem}
\emph{Proof}. Let us consider an arbitrary
$(\widehat{X};\widehat{Y})\in V^2$, then
$$
(\widehat{X};\widehat{Y})=\left(\sum_{l=1}^n c_{1l}
\widehat{v}_{l};\sum_{l=1}^n c_{2l}
\widehat{v}_l\right)=\sum_{l=1}^n (c_{1l}\widehat{v}_l;c_{2l}
\widehat{v}_l),
$$
with $c_{kl}\in \mC(\bo)$ ($k=1,2$). Here the summations in the
second expression is the addition on $V$ and the summation in the
third expression is the addition over $V^2$, i.e. the addition
$+_{V^2}$. Therefore, we have
$$
(\widehat{X};\widehat{Y})=\sum_{l=1}^n c_l \cdot_{V^2}
(\widehat{v}_l;\widehat{v}_l),
$$
where $c_{l}=\eo c_{1l}+\et c_{2l}\in \mathbb{T}$. Moreover, if
$(\widehat{X};\widehat{Y})=(\widehat{0};\widehat{0})$, then
$c_{1l}=c_{2l}=0$ for all $l\in \{1,\ldots,n\}$ since
$\Big\{\widehat{v}_l \mid l\in \{1,\ldots,n\}\Big\}$ is a basis of
$V$ and $c_l=0$ for all $l\in \{1,\ldots,n\}$. Therefore
$\Big\{(\widehat{v}_l;\widehat{v}_l) \mid l\in
\{1,\ldots,n\}\Big\}$ is a $\mathbb{T}$-basis of $V^2$ and the
$\mathbb{T}$-module $(V^2,+_{V^2},\cdot_{V^2})$ is free. It is now
easy to see that $\Big\{\widehat{v}_{l} \mid l\in
\{1,\ldots,n\}\Big\}$ is a $\mathbb{T}$-basis of $M$ since the
isomorphism $\Phi$ given in the proof of Theorem \ref{isoV2M}
gives $\Phi\big((\widehat{v}_l;\widehat{v}_l)\big)=\eo
\widehat{v}_l+\et\widehat{v}_l=\widehat{v}_l$ for all $l\in
\{1,\ldots,n\}$. $\Box$\\

\smallskip\smallskip
\noindent \emph{Remark}. For $(\widehat{X};\widehat{Y})\in V^{2}$,
we have
$$
\begin{array}{rcl}
(\widehat{X};\widehat{Y})&=&(\widehat{X};\widehat{0})+_{V^{2}}(\widehat{0};\widehat{Y})\\*[2ex]
&=&(1\eo +
0\et)\cdot_{V^{2}}(\widehat{X};\widehat{X})+_{V^{2}}(0\eo +
1\et)\cdot _{V^{2}}(\widehat{Y};\widehat{Y})\\*[2ex] &=& \eo\cdot
_{V^{2}}(\widehat{X};\widehat{X})+_{V^{2}} \et\cdot
_{V^{2}}(\widehat{Y};\widehat{Y}),
\end{array}
$$
where $(\widehat{X};\widehat{X})$ and $(\widehat{Y};\widehat{Y})$
are in the vector space $V':=\Big\{\sum_{l=1}^n c_l \cdot_{V^2}
(\widehat{v}_l;\widehat{v}_l)\ |\ c_l\in \mC(\bo)\Big\}$
associated with the free $\mathbb{T}$-module $V^{2}$ using the
following $\mathbb{T}$-basis $\Big\{(\widehat{v}_l;\widehat{v}_l)
\mid l\in \{1,\ldots,n\}\Big\}$.

\smallskip\smallskip\smallskip
\noindent Now, from Theorem \ref{V2basis} we obtain the following corollary.
\begin{corollary}
Let $M$ be a free $\mathbb{T}$-module with a finit
$\mathbb{T}$-basis. The submodule vector space $V$ associated with
$M$ is invariant under a new $\mathbb{T}$-basis of $M$ generated
by another basis of $V$.
\end{corollary}

\section{Bicomplex scalar product}
\label{Bi-scalar}

Let us begin with a preliminary definition.

\begin{definition}
A hyperbolic number $w=a\eo+b\et$ is define to be positive if
$a,b\in \mathbb{R}^{+}$. We denote the set of all positive
hyperbolic numbers by
$$\mathbb{D}^{+}:=\{a\eo+b\et \mid a,b\geq
0\}.
$$
\end{definition}

We are now able to give a definition of a bicomplex scalar
product. (In this article, the physicist convention will be used
for the order of the elements in the bicomplex scalar product.)

\begin{definition}
Let $M$ be a free $\mathbb{T}$-module of finit dimension. With
each pair $\widehat{X}$ and $\widehat{Y}$ in $M$, taken in this
order, we associate a bicomplex number, which is their bicomplex
scalar product $(\widehat{X},\widehat{Y})$, and which satisfies
the following properties:
\\\\
$1.\mbox{ }(\widehat{X},\widehat{Y}_{1}+\widehat{Y}_{2})=(\widehat{X},\widehat{Y}_{1})+(\widehat{X},\widehat{Y}_{2})$,
$\forall \widehat{X},\widehat{Y}_{1},\widehat{Y}_{2}\in M;$\\
$2.\mbox{ }(\widehat{X},\alpha \widehat{Y})=\alpha (\widehat{X},\widehat{Y}),$
$\forall \alpha\in\mathbb{T}$, $\forall \widehat{X},\widehat{Y}\in M;$ \\
$3.\mbox{ }(\widehat{X},\widehat{Y})=(\widehat{Y},\widehat{X})^{\dagger_{3}}$, $\forall \widehat{X},\widehat{Y}\in M;$\\
%$4.\mbox{ }(\widehat{X},\widehat{X})\in\mathbb{D}^{+}$\\
$4.\mbox{ }(\widehat{X},\widehat{X})=0\mbox{
}\Leftrightarrow\mbox{ }\widehat{X}=0$, $\forall \widehat{X}\in
M.$ \label{scalar}
\end{definition}
As a consequence of property~$3$, we have that
$(\widehat{X},\widehat{X})\in\mathbb{D}$.  Note that definition
\ref{scalar} is a general definition of a bicomplex scalar
product. However, in this article we will also require the
bicomplex scalar product $(\cdot,\cdot)$ to be \textit{hyperbolic
positive}, i.e.
\begin{equation}
(\widehat{X},\widehat{X})\in\mathbb{D}^{+},\mbox{
}\forall\widehat{X}\in M \label{hyperpositive}
\end{equation}
and \textit{closed} under the vector space $V$, i.e.
\begin{equation}
(\widehat{X},\widehat{Y})\in\mC(\bo),\mbox{ }\forall
\widehat{X},\widehat{Y}\in V. \label{closed2}
\end{equation}

For the rest of this paper, we will assume a given
$\mathbb{T}$-basis for $M$, which implies a given vector space
$V$.

%So, for the rest of this paper, it will be assumed that $M$ is
%given with the specific $\mathbb{T}$-basis associated with $V$.

\begin{theorem}
Let $\widehat{X},\widehat{Y}\in M$, then
\begin{equation}
(\widehat{X},\widehat{Y})=\eo
(\widehat{X}_{\eo},\widehat{Y}_{\eo})+\et
(\widehat{X}_{\et},\widehat{Y}_{\et}) \label{equ1}
\end{equation}
and
\begin{equation}
P_{k}\big((\widehat{X},\widehat{Y})\big)=(\widehat{X},\widehat{Y})_{\bold{e_k}}=(\widehat{X}_{\bold{e_k}},\widehat{Y}_{\bold{e_k}})\in\mC(\bo)\label{equ2}
\end{equation}
for $k=1,2$.
\label{direct}
\end{theorem}

\noindent \emph{Proof.} From equation (\ref{projection}), it comes
automatically that
$P_{k}\big((\widehat{X},\widehat{Y})\big)=(\widehat{X},\widehat{Y})_{\bold{e_k}}\in\mC(\bo)$
for $k=1,2$. Let $\widehat{X}=\eo \widehat{X}_{\eo}+ \et
\widehat{X}_{\et}$ and $\widehat{Y}=\eo \widehat{Y}_{\eo}+ \et
\widehat{Y}_{\et}$, then using the properties of the bicomplex
scalar product, we also have \bean
(\widehat{X},\widehat{Y}) &=& (\eo \widehat{X}_{\eo}+ \et \widehat{X}_{\et},\eo \widehat{Y}_{\eo}+ \et \widehat{Y}_{\et})\\
      &=& (\eo \widehat{X}_{\eo}+ \et \widehat{X}_{\et},\eo \widehat{Y}_{\eo})+(\eo \widehat{X}_{\eo}+ \et \widehat{X}_{\et},\et \widehat{Y}_{\et})\\
      &=& (\eo \widehat{Y}_{\eo},\eo \widehat{X}_{\eo}+ \et \widehat{X}_{\et})^{\dagger_{3}}+(\et \widehat{Y}_{\et},\eo \widehat{X}_{\eo}+ \et \widehat{X}_{\et})^{\dagger_{3}}\\
      &=& (\eo \widehat{Y}_{\eo},\eo \widehat{X}_{\eo})^{\dagger_{3}}+(\eo \widehat{Y}_{\eo},\et \widehat{X}_{\et})^{\dagger_{3}}\\
      & & +(\et \widehat{Y}_{\et},\eo \widehat{X}_{\eo})^{\dagger_{3}}+(\et \widehat{Y}_{\et},\et \widehat{X}_{\et})^{\dagger_{3}}\\
      &=& \eo^{\dagger_{3}}(\eo \widehat{Y}_{\eo}, \widehat{X}_{\eo})^{\dagger_{3}}+\et^{\dagger_{3}} (\eo \widehat{Y}_{\eo},\widehat{X}_{\et})^{\dagger_{3}}\\
      & & +\eo^{\dagger_{3}}(\et \widehat{Y}_{\et}, \widehat{X}_{\eo})^{\dagger_{3}}+\et^{\dagger_{3}}(\et \widehat{Y}_{\et}, \widehat{X}_{\et})^{\dagger_{3}}\\
      &=& \eo^{\dagger_{3}}\eo (\widehat{X}_{\eo},\widehat{Y}_{\eo})+\et^{\dagger_{3}}\eo (\widehat{X}_{\et},\widehat{Y}_{\eo})\\
      & & +\eo^{\dagger_{3}}\et (\widehat{X}_{\eo},\widehat{Y}_{\et})+\et^{\dagger_{3}}\et (\widehat{X}_{\et},\widehat{Y}_{\et})\\
      &=& \eo (\widehat{X}_{\eo},\widehat{Y}_{\eo})+\et (\widehat{X}_{\et},\widehat{Y}_{\et}).
\eean
Hence,
\begin{equation*}
(\widehat{X},\widehat{Y})=\eo
(\widehat{X}_{\eo},\widehat{Y}_{\eo})+\et
(\widehat{X}_{\et},\widehat{Y}_{\et})
\end{equation*}
and, from property (\ref{closed2}), we obtain
$$P_{k}((\widehat{X},\widehat{Y}))=(\widehat{X},\widehat{Y})_{\bold{e_k}}=(\widehat{X}_{\bold{e_k}},\widehat{Y}_{\bold{e_k}})\in\mC(\bo)$$ for $k=1,2$. $\Box$

\begin{theorem}
$\{V;(\cdot,\cdot)\}$ is a complex $(in\mbox{ }\mC(\bo))$ pre-Hilbert space.
\label{pre-Hilbert}
\end{theorem}

\noindent \emph{Proof.} By definition, $V\subseteq M$.
Hence, we obtain automatically that:

\smallskip
\noindent $1.\mbox{ }(\widehat{X},\widehat{Y}_{1}+\widehat{Y}_{2})=
(\widehat{X},\widehat{Y}_{1})+(\widehat{X},\widehat{Y}_{2})$, $\forall \widehat{X},\widehat{Y}_{1},\widehat{Y}_{2}\in V;$\\
$2.\mbox{ }(\widehat{X},\alpha \widehat{Y})=\alpha
(\widehat{X},\widehat{Y})$, $\forall \alpha\in\mC(\bo)$ and
$\forall \widehat{X},\widehat{Y}\in
V$;\\
$3.\mbox{ }(\widehat{X},\widehat{X})=0\mbox{
}\Leftrightarrow\mbox{ }\widehat{X}=0$, $\forall \widehat{X}\in
V.$
\smallskip

\noindent Moreover, the fact that
$(\widehat{X},\widehat{Y})\in\mC(\bo)$ implies that
$(\widehat{X},\widehat{Y})=(\widehat{Y},\widehat{X})^{\dagger_{3}}=\overline{(\widehat{Y},\widehat{X})}$
and
$(\widehat{X},\widehat{X})\in\mathbb{D}^{+}\cap\mC(\bo)=\mathbb{R}^{+}$.
Hence, $\{V;(\cdot,\cdot)\}$ is a complex $(\mbox{in }\mC(\bo))$
pre-Hilbert space. $\Box$

\smallskip\smallskip
\noindent \emph{Remark}. We note that the results obtained in this
theorem are still valid by using  $\dagger_{1}$ instead of
$\dagger_{3}$ in the definition of the bicomplex scalar product.

\smallskip\smallskip
\noindent Let us denote $\parallel \widehat{X}
\parallel:=(\widehat{X},\widehat{X})^{\frac{1}{2}}$, $\forall \widehat{X}\in V$.

\begin{corollary}
Let $\widehat{X}\in V$. The function $\widehat{X}\longmapsto
\parallel \widehat{X}
\parallel \geq 0$ is a norm on $V$.
\end{corollary}

\begin{corollary}
Let $\widehat{X}\in M$, then
$$P_{k}\big((\widehat{X},\widehat{X})\big)=(\widehat{X},\widehat{X})_{\bold{e_k}}=(\widehat{X}_{\bold{e_k}},\widehat{X}_{\bold{e_k}})=\parallel
\widehat{X}_{\bold{e_k}} \parallel^{2}$$ for $k=1,2$.
\end{corollary}

\noindent Now, let us extend this norm on $M$ with the following
function:
\begin{equation}
\parallel \widehat{X} \parallel:=\Big|(\widehat{X},\widehat{X})^{\frac{1}{2}}\Big|=\Big|\eo \parallel \widehat{X}_{\eo} \parallel+
\et \parallel \widehat{X}_{\et} \parallel\Big|,\mbox{
}\forall{\widehat{X}}\in M. \label{norm}
\end{equation}

\noindent This \textit{norm}  has the following properties.

\begin{theorem}
Let $\widehat{X},\widehat{Y}\in M$ and $d(\widehat{X},\widehat{Y}):=\parallel \widehat{X}-\widehat{Y} \parallel$, then\\
$1.\mbox{ }\parallel \widehat{X} \parallel\geq 0$;\\
$2.\mbox{ }\parallel \widehat{X} \parallel=0\mbox{ }\Leftrightarrow\mbox{ }\widehat{X}=0$;\\
$3.\mbox{ }\parallel \alpha \widehat{X} \parallel=|\alpha|\parallel \widehat{X} \parallel$, $\forall\alpha\in\mC(\bo)$ or $\mC(\bt)$;\\
$4.\mbox{ }\parallel \alpha \widehat{X} \parallel\leq \sqrt{2}\ |\alpha|_{\bold{3}}\parallel \widehat{X} \parallel$, $\forall\alpha\in\mathbb{T}$;\\
$5.\mbox{ }\parallel \widehat{X}+\widehat{Y} \parallel\leq \parallel \widehat{X} \parallel+\parallel \widehat{Y} \parallel$;\\
$6.\mbox{ }\{M,d\}\mbox{ is a \textbf{metric space}.}$\\
\end{theorem}
\emph{Proof.} The proof of $1$ and $2$ come directly from equation
(\ref{norm}). Let $\widehat{X}=\eo \widehat{X}_{\eo}+\et
\widehat{X}_{\et}\in M$ and $\alpha\in\mC(\bo)$ or $\mC(\bt)$,
then \bean
\parallel \alpha \widehat{X} \parallel &=& \Big|(\alpha \widehat{X},\alpha \widehat{X})^{\frac{1}{2}}\Big|\\
                             &=& \left|\big(\alpha\overline{\alpha}(\widehat{X},\widehat{X})\big)^{\frac{1}{2}}\right|\\
                             &=& \left|\left(\eo |\alpha|^{2} (\widehat{X},\widehat{X})_{\eo}+\et |\alpha|^{2} (\widehat{X},\widehat{X})_{\et}\right)^{\frac{1}{2}}\right|\\
                             &=& \Big|\eo |\alpha| (\widehat{X},\widehat{X})_{\eo}^{\frac{1}{2}}+\et |\alpha|(\widehat{X},\widehat{X})_{\et}^{\frac{1}{2}}\Big|\\
                             &=& |\alpha|\, \Big|\eo \parallel \widehat{X}_{\eo} \parallel+\et \parallel \widehat{X}_{\et} \parallel\Big|\\
                             &=& |\alpha|\parallel \widehat{X} \parallel.
\eean
More generally, if $\alpha\in\mathbb{T}$, we obtain
\bean
\parallel \alpha \widehat{X} \parallel &=& \Big|(\alpha \widehat{X},\alpha \widehat{X})^{\frac{1}{2}}\Big|\\
                             &=& \left|\big(\alpha\alpha^{\dagger_{3}}(\widehat{X},\widehat{X})\big)^{\frac{1}{2}}\right|\\
                             &=& \left|\big(|\alpha|_{\bj}^{2}\,(\widehat{X},\widehat{X})\big)^{\frac{1}{2}}\right|\\
                             &=& \Big||\alpha|_{\bj}(\widehat{X},\widehat{X})^{\frac{1}{2}}\Big|\\
                             &=& \Big||\alpha|_{\bj}\,\parallel \widehat{X} \parallel\Big|\\
                             &\leq& \sqrt{2}\,\big||\alpha|_{\bj}\big|\parallel \widehat{X} \parallel\\
                             &=& \sqrt{2}\,|\alpha|_{\bold{3}}\parallel \widehat{X} \parallel.
\eean To complete the proof, we need to establish a triangular
inequality over the $\mathbb{T}$-module $M$. Let
$\widehat{X},\widehat{Y}\in M$, then \bean
\parallel \widehat{X}+\widehat{Y} \parallel &=& |(\widehat{X}+\widehat{Y},\widehat{X}+\widehat{Y})^{\frac{1}{2}}|\\
                        &=& \big|\eo\parallel (\widehat{X}+\widehat{Y})_{\eo} \parallel + \et\parallel (\widehat{X}+\widehat{Y})_{\et} \parallel\big|\\
                        &=& \big|\eo\parallel \widehat{X}_{\eo}+\widehat{Y}_{\eo} \parallel + \et\parallel \widehat{X}_{\et}+\widehat{Y}_{\et} \parallel\big|\\
                        &=& \left( \frac{\parallel \widehat{X}_{\eo}+\widehat{Y}_{\eo} \parallel^{2} + \parallel \widehat{X}_{\et}+\widehat{Y}_{\et} \parallel^{2}}{2}\right)^{\frac{1}{2}}\\
                        &\leq& \left( \frac{\big(\parallel \widehat{X}_{\eo} \parallel +  \parallel \widehat{Y}_{\eo} \parallel\big)^{2} + \big(\parallel \widehat{X}_{\et} \parallel +  \parallel \widehat{Y}_{\et} \parallel\big)^{2}}{2}\right)^{\frac{1}{2}}\\
                        &=& \left|\eo \big(\parallel \widehat{X}_{\eo} \parallel +  \parallel \widehat{Y}_{\eo} \parallel\big)+ \et \big(\parallel \widehat{X}_{\et} \parallel + \parallel \widehat{Y}_{\et} \parallel\big)\right|\\
                        &=& \left|\big(\eo\parallel \widehat{X}_{\eo} \parallel +  \et \parallel \widehat{X}_{\et} \parallel\big)+ \big(\eo\parallel \widehat{Y}_{\eo} \parallel+ \et\parallel \widehat{Y}_{\et} \parallel\big)\right|\\
                        &\leq& \parallel \widehat{X} \parallel+\parallel \widehat{Y} \parallel.
\eean Now, using properties $1$, $2$, $3$ and $5$, it is easy to
obtain that $\{M,d\}$ is a metric space. $\Box$

\smallskip\smallskip\smallskip
With the bicomplex scalar product, it is possible to obtain
a bicomplex version of the well known Schwarz inequality.

\begin{theorem} [Bicomplex Schwarz inequality]
Let $\widehat{X},\widehat{Y}\in M$ then
$$|(\widehat{X},\widehat{Y})|\leq |(\widehat{X},\widehat{X})^{\frac{1}{2}}(\widehat{Y},\widehat{Y})^{\frac{1}{2}}|\leq \sqrt{2}\parallel \widehat{X} \parallel \ \parallel \widehat{Y} \parallel.$$
\end{theorem}

\noindent \emph{Proof.} From the complex (in $\mC(\bo)$) Schwarz
inequality we have that
\begin{equation}
|(\widehat{X},\widehat{Y})|\leq\,\parallel \widehat{X} \parallel \
\parallel \widehat{Y}
\parallel\mbox{ }\forall \widehat{X},\widehat{Y}\in V.
\end{equation}
Therefore, if $\widehat{X},\widehat{Y}\in M$, we obtain \bean
|(\widehat{X},\widehat{Y})| &=& |\eo (\widehat{X},\widehat{Y})_{\eo}+\et (\widehat{X},\widehat{Y})_{\et}|\\
        &=& |\eo (\widehat{X}_{\eo},\widehat{Y}_{\eo})+\et (\widehat{X}_{\et},\widehat{Y}_{\et})|\\
        &=& \left( \frac{|(\widehat{X}_{\eo},\widehat{Y}_{\eo})|^{2}+|(\widehat{X}_{\et},\widehat{Y}_{\et})|^{2}}{2} \right)^{\frac{1}{2}}\\
        &\leq& \left( \frac{\parallel \widehat{X}_{\eo} \parallel^{2} \parallel \widehat{Y}_{\eo} \parallel^{2}+\parallel \widehat{X}_{\et} \parallel^{2} \parallel \widehat{Y}_{\et} \parallel^{2}}{2} \right)^{\frac{1}{2}}\\
        &=& \big|\eo \parallel \widehat{X}_{\eo}\parallel \ \parallel \widehat{Y}_{\eo} \parallel+ \et \parallel \widehat{X}_{\et} \parallel \ \parallel \widehat{Y}_{\et}\parallel\big|\\
        &=& |(\widehat{X},\widehat{X})^{\frac{1}{2}}(\widehat{Y},\widehat{Y})^{\frac{1}{2}}|.
\eean Hence, $|(\widehat{X},\widehat{Y})|\leq
|(\widehat{X},\widehat{X})^{\frac{1}{2}}(\widehat{Y},\widehat{Y})^{\frac{1}{2}}|\leq
\sqrt{2}
\parallel \widehat{X}
\parallel\  \parallel \widehat{Y} \parallel$. $\Box$

\section{Hyperbolic scalar product}

From the preceding section, it is now easy to define the
hyperbolic version of the bicomplex scalar product.

\begin{definition}
Let $M$ be a free $\mathbb{D}$-module of finit dimension. With
each pair $\widehat{X}$ and $\widehat{Y}$ in $M$, taken in this
order, we associate a hyperbolic number, which is their hyperbolic
scalar product $(\widehat{X},\widehat{Y})$, and which satisfies
the following properties:
\\\\
$1.\mbox{ }(\widehat{X},\widehat{Y}_{1}+\widehat{Y}_{2})=(\widehat{X},\widehat{Y}_{1})+(\widehat{X},\widehat{Y}_{2})$; \\
$2.\mbox{ }(\widehat{X},\alpha \widehat{Y})=\alpha
(\widehat{X},\widehat{Y}),$ $\forall \alpha\in\mathbb{D}$; \\
$3.\mbox{ }(\widehat{X},\widehat{Y})=(\widehat{Y},\widehat{X})$;\\
%$4.\mbox{ }(\widehat{X},\widehat{X})\in\mathbb{D}^{+}$\\
$4.\mbox{ }(\widehat{X},\widehat{X})=0\mbox{ }\Leftrightarrow\mbox{ }\widehat{X}=0.$\\
\label{scalar2}
\end{definition}

All definitions and results of Section \ref{Bi-scalar} can be
applied directly in the hyperbolic case if the hyperbolic scalar
product $(\cdot,\cdot)$ is \textit{hyperbolic positive} i.e.
\begin{equation}
(\widehat{X},\widehat{X})\in\mathbb{D}^{+}\mbox{ }\mbox{ }\forall\widehat{X}\in M
\end{equation}
and \textit{closed} under the real vector space
$V:=\left\{\displaystyle \sum_{l=1}^{n}{x_{l}\widehat{m}_l} \mid
x_{l}\in\mathbb{R}\right\}$ i.e.
\begin{equation}
(\widehat{X},\widehat{Y})\in \mC(\bo)\cap \mathbb{D}=\mathbb{R}
\mbox{ }\mbox{ }\forall \widehat{X},\widehat{Y}\in V \label{closed}
\end{equation}
for a specific $\mathbb{D}$-basis $\Big\{\widehat{m}_l \mid l\in
\{1,\ldots,n\}\Big\}$ of $M$. In particular, we obtain an
hyperbolic Schwarz inequality. Moreover, it is always possible to
obtain the angle $\theta$, between $\widehat{X}$ and $\widehat{Y}$
in $V$, with the following well known formula:
\begin{equation}
\cos \theta =\frac{(\widehat{X},\widehat{Y})}{\parallel
\widehat{X}
\parallel\
\parallel \widehat{Y}
\parallel}. \label{angle}
\end{equation}

\noindent From this result, we can derive the following analogue
result for the $\mathbb{D}$-module $M$.

\begin{theorem}
Let $\widehat{X},\widehat{Y}\in M$ and $\theta_{k}$ the angle
between $\widehat{X}_{\bold{e_k}}$ and
$\widehat{Y}_{\bold{e_k}}$ for $k=1,2$. Then,
$$
\cos\left(\frac{\theta_{1}+\theta_{2}}{2}+\frac{\theta_{1}-\theta_{2}}{2}\bj\right)=\frac{(\widehat{X},\widehat{Y})}{(\widehat{X},\widehat{X})^{\frac{1}{2}}(\widehat{Y},\widehat{Y})^{\frac{1}{2}}}.
$$
\end{theorem}

\noindent \emph{Proof.} From the identity (\ref{angle}), we have
\bean
(\cos \theta_{1})\eo+(\cos \theta_{2})\et &=& \frac{(\widehat{X}_{\eo},\widehat{Y}_{\eo})}{\parallel \widehat{X}_{\eo} \parallel\ \parallel \widehat{Y}_{\eo} \parallel}\eo + \frac{(\widehat{X}_{\et},\widehat{Y}_{\et})}{\parallel \widehat{X}_{\et} \parallel\ \parallel \widehat{Y}_{\et} \parallel}\et\\
                                          &=& \frac{(\widehat{X},\widehat{Y})}{(\widehat{X},\widehat{X})^{\frac{1}{2}}(\widehat{Y},\widehat{Y})^{\frac{1}{2}}}.
\eean Moreover, it is easy to show that
$\cos(\theta_{1}\eo+\theta_{2}\et)=(\cos \theta_{1})\eo+(\cos
\theta_{2})\et$ and
$\theta_{1}\eo+\theta_{2}\et=\frac{\theta_{1}+\theta_{2}}{2}+\frac{\theta_{1}-\theta_{2}}{2}\bj$
(see \cite{10}).
Hence, $\cos\left(\frac{\theta_{1}+\theta_{2}}{2}+\frac{\theta_{1}-\theta_{2}}{2}\bj\right)=\frac{(\widehat{X},\widehat{Y})}{(\widehat{X},\widehat{X})^{\frac{1}{2}}(\widehat{Y},\widehat{Y})^{\frac{1}{2}}}$.$\Box$\\

From this result, it is now possible to define the ``hyperbolic
angle'' between two elements of a $\mathbb{D}$-module $M$.

\begin{definition}
Let $\widehat{X},\widehat{Y}\in M$ and $\theta_{k}$ the angle
between $\widehat{X}_{\bold{e_{k}}}$ and $\widehat{Y}_{\bold{e_{k}}}$ for
$k=1,2$. We define the hyperbolic angle between $\widehat{X}$ and
$\widehat{Y}$ as
$$
\frac{\theta_{1}+\theta_{2}}{2}+\frac{\theta_{1}-\theta_{2}}{2}\bj.
$$
\end{definition}

\smallskip\smallskip
\noindent We note that our definition of the hyperbolic scalar
product is different from the definitions given in \cite{7, 8} and
\cite{12}.

\section{Bicomplex Hilbert space}

\begin{definition}
Let $M$ be a free  $\mathbb{T}$-module with a finit
$\mathbb{T}$-basis. Let also $(\cdot,\cdot)$ be a bicomplex scalar
product defined on $M$. The space $\{M, (\cdot,\cdot)\}$ is called
a $\mathbb{T}$-inner product space.
\end{definition}

\begin{definition}
A complete $\mathbb{T}$-inner product space is called a $\mathbb{T}$-Hilbert space.
\end{definition}

\begin{lemma}
Let $\widehat{X}\in M$ then
$$\parallel \widehat{X}_{\bold{e_k}} \parallel\leq\sqrt{2} \parallel \widehat{X} \parallel,\mbox{ for } k=1,2.$$
\label{projection2}
\end{lemma}
\emph{Proof.} For $k=1,2$, we have \bean
\parallel \widehat{X}_{\bold{e_k}}\parallel &\leq& \sqrt{2}\left(\frac{\parallel \widehat{X}_{\mathrm{e}_{1}}\parallel^{2}+\parallel \widehat{X}_{\mathrm{e}_{2}}\parallel^{2}}{2}\right)^{\frac{1}{2}}\\
                             &=& \sqrt{2}\,\big| \eo \parallel \widehat{X}_{\eo}\parallel + \et \parallel \widehat{X}_{\et}\parallel\big|\\
                             &=& \sqrt{2} \parallel \widehat{X} \parallel.\ \Box
\eean

\begin{lemma}
The pre-Hilbert space $\{V, (\cdot,\cdot)\}$ is closed in the
metric space $\{M, (\cdot,\cdot)\}$.
\end{lemma}
\emph{Proof.} Let $\widehat{X}_{n}=\eo \widehat{X}_{n} + \et
\widehat{X}_{n}\in V\mbox{ }\forall n\in \mathbb{N}$ and
$\widehat{X}=\eo \widehat{X}_{\eo} + \et \widehat{X}_{\et}\in M$.
Supposed that $\widehat{X}_n\rightarrow \widehat{X}$ whenever $n\rightarrow
\infty$ then $\parallel \widehat{X}_n-\widehat{X}
\parallel\rightarrow 0$ as $n\rightarrow \infty$ i.e.
$\parallel \widehat{X}_{n} - (\eo \widehat{X}_{\eo} + \et
\widehat{X}_{\et})\parallel$= $\parallel (\eo \widehat{X}_{n} +
\et \widehat{X}_{n} ) - (\eo \widehat{X}_{\eo} + \et
\widehat{X}_{\et})\parallel$ =$\parallel \eo(\widehat{X}_{n} -
\widehat{X}_{\eo}) + \et(\widehat{X}_{n} - \widehat{X}_{\et})
\parallel\rightarrow 0$ as $n\rightarrow \infty$. Therefore, from the
Lemma \ref{projection2} we have that \bean
\parallel  \widehat{X}_{n} - \widehat{X}_{\bold{e_k}} \parallel &\leq& \sqrt{2}\parallel \eo(\widehat{X}_{n} - \widehat{X}_{\eo}) + \et(\widehat{X}_{n} - \widehat{X}_{\et}) \parallel\rightarrow 0\\
\eean as $n\rightarrow \infty$ for $k=1,2$. Hence,
$\widehat{X}_{\eo}=\widehat{X}_{\et}=\widehat{X}$ and
$\widehat{X}=\eo \widehat{X} + \et \widehat{X}\in V$. $\Box$

\begin{theorem}
A $\mathbb{T}$-inner product space $\{M, (\cdot,\cdot)\}$ is a
$\mathbb{T}$-Hilbert space if and only if $\{V, (\cdot,\cdot)\}$
is an Hilbert space. \label{Hilbert}
\end{theorem}

\noindent \emph{Proof.} From the Theorem \ref{pre-Hilbert}, $\{V,
(\cdot,\cdot)\}$ is a pre-Hilbert space. So, we have to prove that
$\{M, (\cdot,\cdot)\}$ is complete if and only if $\{V,
(\cdot,\cdot)\}$ is complete. By definition $V\subseteq M$,
therefore if $M$ is complete then $V$ is also complete since $V$ is
closed in $M$. Conversely, let $\widehat{X}_{n}=\eo
(\widehat{X}_{n})_{\eo} + \et (\widehat{X}_{n})_{\et}\in M\mbox{
}\forall n\in \mathbb{N}$, be a Cauchy sequence in $M$. Then, from
the Lemma \ref{projection2}, we have \bean
\parallel (\widehat{X}_{m})_{\bold{e_k}} - (\widehat{X}_{n})_{\bold{e_k}} \parallel = \parallel  (\widehat{X}_{m} - \widehat{X}_{n})_{\bold{e_k}} \parallel \leq \sqrt{2}\parallel \widehat{X}_{m} - \widehat{X}_{n} \parallel
\eean for $k=1,2$. So, $(\widehat{X}_{n})_{\bold{e_k}}$ is also
a Cauchy sequence in $V$ for $k=1,2$. Therefore, there exist
$\widehat{X}_{\eo},\widehat{X}_{\et}\in V$ such that
$(\widehat{X}_{n})_{\bold{e_k}}\rightarrow
\widehat{X}_{\bold{e_k}}$ as $n\rightarrow \infty$ for $k=1,2$.

Now, from the triangular inequality, if we let $\widehat{X}:=\eo
\widehat{X}_{\eo} + \et \widehat{X}_{\et}$, then we obtain
 \bean
\parallel \widehat{X}_{n}-\widehat{X} \parallel &=& \parallel \eo \big((\widehat{X}_{n})_{\eo}-\widehat{X}_{\eo}\big) + \et \big((\widehat{X}_{n})_{\et}-\widehat{X}_{\et}\big)\parallel\\
                            &\leq& \parallel \eo \big((\widehat{X}_{n})_{\eo}-\widehat{X}_{\eo}\big)  \parallel + \parallel \et \big((\widehat{X}_{n})_{\et}-\widehat{X}_{\et}\big)\parallel\\
                            &\leq& \sqrt{2}\,|\eo|_{\bold{3}}\parallel  (\widehat{X}_{n})_{\eo}-\widehat{X}_{\eo}  \parallel\\
                            & & +\sqrt{2}\,|\et|_{\bold{3}}\parallel  (\widehat{X}_{n})_{\et}-\widehat{X}_{\et}\parallel\\
                            &=& \parallel (\widehat{X}_{n})_{\eo}-\widehat{X}_{\eo}  \parallel + \parallel (\widehat{X}_{n})_{\et}-\widehat{X}_{\et}\parallel\rightarrow 0
\eean as $n\rightarrow \infty$. Hence, $\widehat{X}_{n}\rightarrow
\widehat{X}\in M$ as $n\rightarrow \infty$. $\Box$

\subsection*{Examples of bicomplex Hilbert spaces}
\begin{enumerate}
\item Let us first consider $M=\mathbb{T}$, the canonical
$\mathbb{T}$-module over the ring of bicomplex numbers.
We consider now the trivial $\mathbb{T}$-basis $\{1\}$. In this
case, the submodule vector space $V$ is simply $V=\mC(\bo)$. Let
$(\cdot,\cdot)_{1}$ and $(\cdot,\cdot)_{2}$ be two scalar product
on $V$. It is always possible to construct a general bicomplex
scalar product as follows:

\smallskip
Let $$w_1=(z_{11}-z_{12}\bo)\eo+(z_{11}+z_{12}\bo)\et$$
and
$$w_2=(z_{21}-z_{22}\bo)\eo+(z_{21}+z_{22}\bo)\et $$ where,
$z_{11},z_{12},z_{21},z_{22}\in \mC(\bo)$. We define
\begin{equation}
(w_1,w_2):=(z_{11}-z_{12}\bo,z_{21}-z_{22}\bo)_1\eo+(z_{11}+z_{12}\bo,z_{21}+z_{22}\bo)_2\et.
\end{equation}
However, this bicomplex scalar product is not \textit{closed}
under $\mC(\bo)$. In fact, $(\cdot,\cdot)$ will be \textit{closed}
under $\mC(\bo)$ if and only if
$(\cdot,\cdot)_{1}=(\cdot,\cdot)_{2}$. From the Theorem
\ref{Hilbert}, we obtain the following result.

\begin{theorem}
Let $\mathbb{T}$, the canonical  $\mathbb{T}$-module over the ring
of bicomplex numbers with a scalar product $(\cdot,\cdot)$ on
$\mC(\bo)$. Let also
$w_1=(z_{11}-z_{12}\bo)\eo+(z_{11}+z_{12}\bo)\et$ and
$w_2=(z_{21}-z_{22}\bo)\eo+(z_{21}+z_{22}\bo)\et $ where,
$z_{11},z_{12},z_{21},z_{22}\in \mC(\bo)$. If we define
\begin{equation}
(w_1,w_2):=(z_{11}-z_{12}\bo,z_{21}-z_{22}\bo)\eo+(z_{11}+z_{12}\bo,z_{21}+z_{22}\bo)\et
\end{equation}
then $\{\mathbb{T},(\cdot,\cdot)\}$ is a bicomplex Hilbert space
if and only if $\{\mC(\bo),(\cdot,\cdot)\}$ is an Hilbert space.
\label{za}
\end{theorem}

\smallskip
As an example, let us consider $\{\mC(\bo),(\cdot,\cdot)\}$ with
the canonical scalar product given by
\bean
(z_1,z_2) &=& (x_1+y_1\bo,x_2+y_2\bo)\\
          &:=& x_1x_2+y_1y_2.
\eean It is well known that $\{\mC(\bo),(\cdot,\cdot)\}$ is an
Hilbert space. Hence, from the Theorem \ref{za},
$\{\mathbb{T},(\cdot,\cdot)\}$ is a bicomplex Hilbert space.
Moreover, it is easy to see that
$$\parallel w \parallel=||w|_{\bj}|=|w|_{\bold{3}}=|w|,$$
i.e. the Euclidean metric of $\mathbb{R}^{4}$.

\newpage
\item Consider now $M=\mathbb{T}^n$, the $n$-dimensional module with the canonical
$\mathbb{T}$-basis $\{\widehat{e}_i\ |\ i\in\{1,\ldots,n\}\}$, the
columns of the identity matrix $I_n$. For any two elements
$\widehat{X},\widehat{Y}\in \mathbb{T}^n$ given by
$\widehat{X}=\displaystyle \sum_{i=1}^{n} x_i\,\widehat{e}_i$ and
$\widehat{Y}=\displaystyle \sum_{i=1}^{n} y_i\,\widehat{e}_i$, we
define the bicomplex scalar product as
\begin{equation}
(\widehat{X},\widehat{Y}):=(\widehat{X}^{\dag_3})^\top \cdot
\widehat{Y}=\displaystyle \sum_{i=1}^{n}x_{i}^{\dag_3}\,y_i \in
\mathbb{T}. \label{scalarproductTn}
\end{equation}
It is now easy to verify that properties $1$, $2$ and $3$ of
Definition \ref{scalar} are trivially satisfied. This bicomplex
scalar product also implies that $(\widehat{X},\widehat{X})=
\sum_{i=1}^{n}x_{i}^{\dag_3}\,x_i=\sum_{i=1}^{n}|x_{i}|_{\bj}^2=
\eo \sum_{i=1}^{n}|x_{1i}-x_{2i}\bo|^2 + \et \sum_{i=1}^{n}|x_{1i}+x_{2i}\bo|^2$
where $x_i=x_{1i}+x_{2i}\bt=(x_{1i}-x_{2i}\bo)\eo +(x_{1i}+x_{2i}\bo)\et$ for $i\in\{1,\ldots,n\}$.
Hence, the property $4$ of Definition \ref{scalar} is also satisfied and
\begin{equation}
\parallel \widehat{X}\parallel =|(\widehat{X},\widehat{X})^{\frac{1}{2}}|=\Big|\big(\sum_{i=1}^{n}|x_{i}|_{\bj}^2\big)^{\frac{1}{2}}\Big|.
\label{prodscalXXTn}
\end{equation}
In this example, the complex vector space $V=\{\sum_{i=1}^{n}x_i
\widehat{e}_i \ |\ x_i\in \mathbb{C}(\bo)\}$ is simply the
standard complex vector space isomorphic to $\mathbb{C}^n$.
Moreover, the closure property is satisfied since for
$\widehat{X},\widehat{Y}\in V$ we have $x_i,y_i\in
\mathbb{C}(\bo)$ and $x_{i}^{\dag_3}\,y_i=\overline{x}_{i}\,y_i\in
\mathbb{C}(\bo)$ such that equation (\ref{scalarproductTn}) gives
an element of $\mathbb{C}(\bo)$.

\end{enumerate}

\section{The Dirac notation over $M$}
%\subsection{Bicomplex scalar product: Ket}
In this section we introduce the Dirac notation usually used in
quantum mechanics. For this we have to define correctly kets and
bras over a bicomplex Hilbert space which, we remind, is
fundamentally a module.

Let $M$ be a  $\mathbb{T}$-module which is free with the following
finit $\mathbb{T}$-basis $\{\ket{m_l} \mid l\in \{1,\ldots,n\}\}$.
Any element of $M$ will be called a \textit{ket module} or, more
simply, a \textit{ket}.
%It is
%represented by the symbole $|\cdot>$, inside which is placed a distinctive sign
%which enable us to distinguish the corresponding ket from all others, for example:
%$|\psi>$.

Let us rewrite the definition of the bicomplex scalar product in
term of the ket notation.

\begin{definition}
Let $M$ be a  $\mathbb{T}$-module which is free with the following
finit $\mathbb{T}$-basis $\{\ket{m_l} \mid l\in \{1,\ldots,n\}\}$.
With each pair $\ket{\phi}$ and $\ket{\psi}$ in $M$, taken in this
order, we associate a bicomplex number, which is their bicomplex
scalar product $(\ket{\phi},\ket{\psi})$, and which satisfies the
following properties:
\\\\
$1.\mbox{ }(\ket{\phi},\ket{\psi_1}+\ket{\psi_2})=(\ket{\phi},\ket{\psi_1})+(\ket{\phi},\ket{\psi_2})$;\\
$2.\mbox{ }(\ket{\phi},\alpha \ket{\psi})=\alpha (\ket{\phi},\ket{\psi}),$ $\forall \alpha\in\mathbb{T}$;\\
$3.\mbox{ }(\ket{\phi},\ket{\psi})=(\ket{\psi},\ket{\phi})^{\dagger_{3}}$;\\
%$4.\mbox{ }(\ket{\phi},\ket{\phi})\in\mathbb{D}^{+}$\\
$4.\mbox{ }(\ket{\phi},\ket{\phi})=0\mbox{ }\Leftrightarrow\mbox{
}\ket{\phi}=0.$ \label{scalar+ket}
\end{definition}

%\subsection{Element of the dual space $M^{\ast}$ of $M$: Bras}
\noindent Let us now define the dual space $M^{\ast}$.

\begin{definition}
A linear functional $\chi$ is a linear operation which associates
a bicomplex number with every ket $\ket{\psi}$:

\smallskip\smallskip
\noindent $1)$ $\ket{\psi}\longrightarrow
\chi(\ket{\psi})\in\mathbb{T}$;

\noindent $2)$
$\chi(\lambda_1\ket{\psi_1}+\lambda_2\ket{\psi_2})=\lambda_1\chi(\ket{\psi_1})+\lambda_2\chi(\ket{\psi_2}),\
\ \lambda_1,\lambda_2\in\mathbb{T}.$
\smallskip

\noindent It can be shown that the set of linear functionals
defined on the kets $\ket{\psi}\in M$ constitutes a
$\mathbb{T}$-module space, which is called the dual space of $M$
and which will be symbolized by $M^{\ast}$.
\end{definition}

\noindent Using this definition of $M^{\ast}$, let us define the
bra notation.

\begin{definition}
Any element of the space $M^{\ast}$ is called a bra module or,
more simply, a bra. It is symbolized by $\bra{\,\cdot\,}$.
\end{definition}

For example, the bra $\bra{\chi}$ designates the bicomplex linear
functional $\chi$ and we shall henceforth use the notation
$\braket{\chi}{\psi}$ to denote the number obtained by causing the
linear functional $\bra{\chi}\in M^{\ast}$ to act on the ket
$\ket{\psi}\in M$:
$$\chi(\ket{\psi}):=\braket{\chi}{\psi}.$$

%\subsection{Correspondence between kets and bras}

The existence of a bicomplex scalar product in $M$ will now enable
us to show that we can associate, with every ket $\ket{\phi}\in
M$, an element of $M^{\ast}$, which will be denoted by
$\bra{\phi}$.

The ket $\ket{\phi}$ does indeed enable us to define a linear
functional: the one which associates (in a linear way), with each
ket $\ket{\psi}\in M$, a bicomplex numbers which is equal to the
scalar product $(\ket{\phi},\ket{\psi})$ of $\ket{\psi}$ by
$\ket{\phi}$. Let $\bra{\phi}$ be this linear functional; It is
thus defined by the relation: \be
\braket{\phi}{\psi}=(\ket{\phi},\ket{\psi}).
\label{linearfunctdef}\ee

\noindent Therefore, the properties of the bicomplex scalar product can be
rewrited as:
\smallskip\smallskip

\noindent $1.\mbox{ }\bra{\phi}\big(\ket{\psi_{1}}+\ket{\psi_{2}}\big)=\braket{\phi}{\psi_1}+\braket{\phi}{\psi_2}$;\\
\noindent $2.\mbox{ }\braket{\phi}{\alpha \psi}=\alpha\,\braket{\phi}{\psi},$ $\forall \alpha\in\mathbb{T}$;\\
\noindent $3.\mbox{ }\braket{\phi}{\psi}=\braket{\psi}{\phi}^{\dagger_{3}}$;\\
%\noindent $4.\mbox{ }\braket{\phi}{\phi}\in\mathbb{D}^{+}$\\
\noindent $4.\mbox{ }\braket{\phi}{\phi}=0\mbox{ }\Leftrightarrow\mbox{ }\ket{\phi}=0.$\\

Now, let define the corresponding projections for the Dirac notation as
follows.

\begin{definition}
Let $\ket{\psi}$,$\ket{\phi}\in M$ and $\ket{\chi}\in V$. For
$k=1,2$, we define:

\smallskip\smallskip
%\noindent $1.$ $\braket{\phi}{\psi}_{\bold{e_k}}:=P_{k}(\braket{\phi}{\psi})$\\
\noindent $1.$ $\ket{\psi_{\bold{e_k}}}:=P_{k}(\ket{\psi})\in V$;\\
\noindent $2.$
$\bra{\phi_{\bold{e_k}}}:=P_{k}(\bra{\phi}):V\longrightarrow\mC(\bo)$,
where $\ket{\chi}\mapsto P_{k}\big(\braket{\phi}{\chi}\big).$
\end{definition}
The first definition gives the projection
$\ket{\psi_{\bold{e_k}}}$ of the ket $\ket{\psi}$ of $M$. This
is well defined from equation (\ref{projection}). However, the
second definition is more subtle. In the next two theorems, we
show that $\bra{\phi_{\bold{e_k}}}$ is really the bra
associated with the ket $\ket{\phi_{\bold{e_k}}}$ in $V$.

\begin{theorem}
Let $\ket{\phi}\in M$, then
$$\bra{\phi_{\bold{e_k}}}\in V^{\ast}$$ for $k=1,2$.
\end{theorem}

\noindent \emph{Proof.} Let $\lambda_1,\lambda_2\in\mC(\bo)$ and
$\ket{\psi_1},\ket{\psi_2}\in V$, then
$$
\begin{array}{rcl}
\bra{\phi_{\bold{e_k}}}(\lambda_1\ket{\psi_{1}}+\lambda_2\ket{\psi_{2}})
&=&
P_k\Big(\bra{\phi}\big(\lambda_1\ket{\psi_{1}}+\lambda_2\ket{\psi_{2}}\big)\Big)\\*[2ex]
&=& P_k\Big(\lambda_1 \braket{\phi}{\psi_{1}}+\lambda_2
\braket{\phi}{\psi_{2}}\Big)\\*[2ex] &=& \lambda_1
P_k\Big(\braket{\phi}{\psi_{1}}\Big)+\lambda_2
P_k\Big(\braket{\phi}{\psi_{2}}\Big)\\*[2ex] &=&\lambda_1
\bra{\phi_{\bold{e_k}}}(\ket{\psi_{1}})+\lambda_2
\bra{\phi_{\bold{e_k}}}(\ket{\psi_{2}})
\end{array}
$$
for $k=1,2$. $\Box$\\

\noindent We will now show that the functional
$\bra{\phi_{\bold{e_k}}}$ can be obtained from the ket
$\ket{\phi_{\bold{e_k}}}$.

\begin{theorem}
Let $\ket{\phi}\in M$ and $\ket{\psi}\in V$, then
\begin{equation}
\bra{\phi_{\bold{e_k}}}(\ket{\psi})=\braket{\phi_{\bold{e_k}}}{\psi}
\end{equation}
for $k=1,2$.
\label{direct2}
\end{theorem}

\noindent \emph{Proof.}
Using (\ref{equ2}) in Theorem \ref{direct} and the fact that $P_k(\ket{\psi})=\ket{\psi}$, we obtain
$$
\begin{array}{rcl}
\bra{\phi_{\bold{e_k}}}(\ket{\psi}) &=& P_k\Big(\braket{\phi}{\psi}\Big)\\
                                        &=& P_k\Big((\ket{\phi},\ket{\psi})\Big)\\
                                        &=& \Big(P_k(\ket{\phi}),P_k(\ket{\psi})\Big)\\
                                        &=& \Big(P_k(\ket{\phi}),\ket{\psi}\Big)\\
                                        &=& \Big(\ket{\phi_{\bold{e_k}}},\ket{\psi}\Big)\\
                                        &=& \braket{\phi_{\bold{e_k}}}{\psi}
\end{array}
$$
for $k=1,2$. $\Box$\\

\begin{corollary}
Let $\ket{\phi},\ket{\psi}\in M$ then
\begin{equation}
\braket{\phi_{\bold{e_k}}}{\psi_{\bold{e_k}}}=\braket{\phi}{\psi}_{\bold{e_k}}
\end{equation}
for $k=1,2$.
\label{deco}
\end{corollary}
\emph{Proof}. From Theorem \ref{direct2} and the properties of the
projectors $P_k$, we obtain
$$
\begin{array}{rcl}
\braket{\phi_{\bold{e_k}}}{\psi_{\bold{e_k}}} &=& P_k\Big(\braket{\phi}{\psi_{\bold{e_k}}}\Big)\\
                                                      &=& P_k\Big(\eo \braket{\phi}{\psi_{\eo}}+\et\braket{\phi}{\psi_{\et}}\Big)\\
                                                      &=& P_k\Big(\bra{\phi}(\eo \ket{\psi_{\eo}}+ \et \ket{\psi_{\et}})\Big)\\
                                                      &=& P_k\Big(\braket{\phi}{\psi}\Big)\\
                                                      &=& \braket{\phi}{\psi}_{\bold{e_k}}
\end{array}
$$
for $k=1,2$. $\Box$\\

The bicomplex scalar product is antilinear. Indeed, by using the
notation (\ref{linearfunctdef}) we obtain
$$
\begin{array}{rcl}
(\lambda_1\ket{\phi_1}+\lambda_2 \ket{\phi_2},\
\ket{\psi})&=&(\ket{\psi},\ \lambda_1\ket{\phi_1}+\lambda_2
\ket{\phi_2})^{\dag_3} \\*[2ex] &=& (\lambda_1
\braket{\psi}{\phi_1}+\lambda_2 \braket{\psi}{\phi_2})^{\dag_3}
\\*[2ex]
&=&
\lambda_1^{\dag_3}\braket{\phi_1}{\psi}+\lambda_2^{\dag_3}\braket{\phi_2}{\psi}
\\*[2ex]
&=&
\big(\lambda_1^{\dag_3}\bra{\phi_1}+\lambda_2^{\dag_3}\bra{\phi_2}\big)\ket{\psi},
\end{array}
$$
where $\lambda_1,\lambda_2 \in \mathbb{T}$ and
$\ket{\psi},\ket{\phi_1},\ket{\phi_2}\in M$. Therefore the bra
associated with the ket $\lambda_1\ket{\phi_1}+\lambda_2
\ket{\phi_2}$ is given by
$\lambda_1^{\dag_3}\bra{\phi_1}+\lambda_2^{\dag_3}\bra{\phi_2}$:
$$
\lambda_1\ket{\phi_1}+\lambda_2 \ket{\phi_2} \leftrightsquigarrow
\lambda_1^{\dag_3}\bra{\phi_1}+\lambda_2^{\dag_3}\bra{\phi_2}.
$$
In particular, Theorem \ref{theo:Xe1+Xe2} tell us that every ket
$\ket{\psi}\in M$ can be written in the form $\ket{\psi}=\eo
\ket{\psi_{\eo}}+\et \ket{\psi_{\et}}$. Therefore, we have $
\ket{\psi}=\eo \ket{\psi_{\eo}}+\et \ket{\psi_{\et}}
\leftrightsquigarrow \bra{\psi}=\eo \bra{\psi_{\eo}}+\et
\bra{\psi_{\et}} $ since $(\bold{e_k})^{\dag_3}=\bold{e_k}$
for $k=1,2$.

\section{Bicomplex linear operators}
\subsection{Basic results and definitions}
The \emph{bicomplex linear operators} $A: M\rightarrow M$ are
defined by
$$
\begin{array}{l}
\ket{\psi'}=A\ket{\psi}, \\*[2ex]
A(\lambda_1\ket{\psi_1}+\lambda_2\ket{\psi_2})=\lambda_1
A\ket{\psi_1}+\lambda_2 A\ket{\psi_2},
\end{array}
$$
where $\lambda_1,\lambda_2\in \mathbb{T}$. For a fixed
$\ket{\phi}\in M$, a fixed linear operator $A$ and an arbitrary
$\ket{\psi}\in M$, we define the bra $\bra{\phi}A$ by the relation
$$
\big(\bra{\phi}A\big)\ket{\psi}:=\bra{\phi}\big(A\ket{\psi}\big).
$$
The operator $A$ associates a new bra $\bra{\phi}A$ for every bra
$\bra{\phi}$. It is easy to show that this correspondance is
linear, i.e. $(\lambda_1 \bra{\phi_1}+\lambda_2
\bra{\phi_2})A=\lambda_1 \bra{\phi_1}A+\lambda_2 \bra{\phi_2}A$.

For a given linear operator $A:M\rightarrow M$, the
\emph{bicomplex adjoint operator} $A^*$ is the operator with the
following correspondance
\begin{equation}
\ket{\psi'}=A\ket{\psi} \leftrightsquigarrow
\bra{\psi'}=\bra{\psi} A^*. \label{defopad}
\end{equation}
The bicomplex adjoint operator $A^*$ is linear: the proof is
analogous to the standard case except that the standard complex
conjugate is replaced by $\dag_3$ everywhere. Note that since we
have $\braket{\psi'}{\phi}=\braket{\phi}{\psi'}^{\dag_3}$, we
obtain
\begin{equation}
\bra{\psi}A^*\ket{\phi}=\bra{\phi}A\ket{\psi}^{\dag_3},
\end{equation}
by using expressions (\ref{defopad}).

It is easy to show that for any bicomplex linear operator
$A:M\rightarrow M$ and $\lambda\in \mathbb{T}$, we have the
following standard properties:

\begin{eqnarray}
(A^*)^*&=&A, \\
\label{lAe}
(\lambda A)^*&=&\lambda^{\dag_3}A^*, \\
(A+B)^*&=&A^*+B^*, \\
(AB)^*&=&B^*A^*.
\end{eqnarray}
These properties are prove similarly as the standard cases.

\begin{definition}
Let $M$ be a bicomplex Hilbert space and  $A:M\rightarrow M$ a
bicomplex linear operator. We define the projection $P_k(A):M\rightarrow V$ of $A$,
for $k=1,2$, as follows :
$$
P_k(A)\ket{\psi}:=P_k(A\ket{\psi}),\mbox{ }\mbox{ }\forall\,
\ket{\psi}\in M.
$$
\label{ProA}
\end{definition}

\noindent The projection $P_k(A)$ is clearly a bicomplex linear operator for $k=1,2$. Moreover,
we have the following specific results.

\begin{theorem}
Let $M$ be a bicomplex Hilbert space, $A:M\rightarrow M$ a
bicomplex linear operator and $\ket{\psi}=\eo \ket{\psi_{\eo}}+\et
\ket{\psi_{\et}}\in M$. Then
\begin{enumerate}
    \item[$(i)$] $A\ket{\psi}=\eo P_1(A)\ket{\psi_{\eo}} +\et P_2(A)\ket{\psi_{\et}}$;
    \item[$(ii)$] $P_k(A)^{*}=P_k({A^*})$ where $P_k(A)^{*}$ is the standard complex adjoint operator over $\mathbb{C}(\bo)$ associated with
the bicomplex linear operator $P_k(A)$ restricted to the submodule vector space $V$, defined in
$(\ref{V})$, for $k=1,2.$
\end{enumerate}
\label{AcloseV}
\end{theorem}

\noindent \emph{Proof.} The part (\emph{i}) is obtain as follows:
$$
\begin{array}{rcl}
A\ket{\psi} &=& A\big(\eo \ket{\psi_{\eo}}+\et \ket{\psi_{\et}}\big)\\
            &=& \eo A\ket{\psi_{\eo}} + \et A\ket{\psi_{\et}}\\

            &=& \eo \Big(\eo P_1(A\ket{\psi_{\eo}}) + \et P_2(A\ket{\psi_{\eo}})\Big)\\
            & & + \et \Big(\eo P_1(A\ket{\psi_{\et}}) + \et P_2(A\ket{\psi_{\et}})\Big)\\

            &=& \eo \Big(\eo P_1(A)\ket{\psi_{\eo}} + \et P_2(A)\ket{\psi_{\eo}}\Big)\\
            & & + \et \Big(\eo P_1(A)\ket{\psi_{\et}} + \et P_2(A)\ket{\psi_{\et}}\Big)\\
            &=& \eo P_1(A)\ket{\psi_{\eo}} +\et P_2(A)\ket{\psi_{\et}}.
\end{array}
$$
To show (\emph{ii}), we use (\emph{i}) and Corollary \ref{deco} to decompose the
correspondance (\ref{defopad}) into the equivalent following
correspondance in $V$:
\begin{equation}
\ket{\psi_{\bold{e_k}}'}=P_k(A)\ket{\psi_{\bold{e_k}}} \leftrightsquigarrow
\bra{\psi_{\bold{e_k}}'}=\bra{\psi_{\bold{e_k}}} P_k(A^*)\mbox{ for }k=1,2. \label{defopad2}
\end{equation}
Hence, $P_k(A)^{*}=P_k({A^*})$.$\Box$

\subsection{Bicomplex eigenvectors and eigenvalues on $M$}

One can show now that the bicomplex eigenvector equation
$A\ket{\psi}=\lambda \ket{\psi}$, with $\lambda\in \mathbb{T}$, is
equivalent to the system of two eigenvector equations given by
$$
\begin{array}{rcl}
P_1(A)\ket{\psi_{\eo}}&=&\lambda_1 \ket{\psi_{\eo}}, \\
P_2(A)\ket{\psi_{\et}}&=&\lambda_2 \ket{\psi_{\et}},
\end{array}
$$
where $\lambda=\eo \lambda_1+\et \lambda_2$,
$\lambda_1,\lambda_2\in \mathbb{C}(\bo)$ and $\ket{\psi}=\eo
\ket{\psi_{\eo}}+\et \ket{\psi_{\et}}$. Indeed, we have
\begin{eqnarray}
A\ket{\psi}=\lambda \ket{\psi}& \Leftrightarrow & A\ket{\psi}=(\lambda_1\eo+\lambda_2\et)(\eo
\ket{\psi_{\eo}}+\et \ket{\psi_{\et}}) \nonumber \\*[2ex] &
\Leftrightarrow & \eo P_1(A)\ket{\psi_{\eo}}+\et P_2(A)
\ket{\psi_{\et}}=\eo\lambda_1 \ket{\psi_{\eo}}+\et\lambda_2
\ket{\psi_{\et}} \nonumber \\*[2ex] & \Leftrightarrow &
P_k(A)\ket{\psi_{\bold{e_k}}}=\lambda_k \ket{\psi_{\bold{e_k}}},\ \
\ \ k=1,2.\ \label{syseigen}
\end{eqnarray}

Suppose now that $\Big\{\ket{v_l}\ |\ l\in\{1,\ldots,n\}\Big\}$ is
an orthonormal basis of $V$ (which is also a basis of $M$ from
Theorem \ref{V2basis}) with
$\ket{\psi_{\bold{e_k}}}=\displaystyle \sum_{j=1}^n
c_{kj}\ket{v_j}$, $c_{kj}\in \mathbb{C}(\bo)$, $k=1,2$. Then from
(\ref{syseigen}) we find $\displaystyle \sum_{j=1}^n c_{kj}
P_k(A)\ket{v_j}=\lambda_k \sum_{j=1}^n c_{kj}\ket{v_j}$ for
$k=1,2$. Applying now the functional $\bra{v_i}$ on this
expression, we obtain
$$
\begin{array}{rcl}
\displaystyle \sum_{j=1}^n c_{kj} \bra{v_i}P_k(A)\ket{v_j}&=&\lambda_k
\displaystyle \sum_{j=1}^n c_{kj}\braket{v_i}{v_j}\\*[2ex]
&=&\lambda_k c_{ki},
\end{array}
$$
where the last line is a consequence of the orthogonality
$\braket{v_i}{v_j}=\delta_{ij}$ of the basis of $V$. Now, by
definition, we have that $P_k(A)\ket{v_j}\in V$ for $k=1,2$.
Moreover since $\ket{v_i}$ is also an element of $V$ then the
closure of the scalar product of two elements of $V$, see equation
(\ref{closed2}), implies that the matrix $A_k$ defined by
$$
\begin{array}{l}
(A_k)_{ij}:=\bra{v_i}P_k(A)\ket{v_j}
\end{array}
$$
is in $\mathbb{C}(\bo)$ for $k=1,2$. Therefore, we find that
$$
\sum_{j=1}^{n} \big((A_k)_{ij}-\lambda_k \delta_{ij}\big)c_{kj}=0,
\ \ \ k=1,2.
$$
Each equation, i.e. $k=1$ and $k=2$, is a homogeneous linear
system with $n$ equations and $n$ unknowns which can be solved
completely since all components are in $\mathbb{C}(\bo)$.
Therefore, the system posses a nontrivial solution if and only
$\det(A_k-\lambda_k I_n)=0$ for $k=1,2$.

In standard quantum mechanics self-adjoint operators (Hermitian
operators) play a very important role. In analogy with the
standard case, a linear operator $A$ is defined to be a
\emph{bicomplex self-adjoint operator} if and only if $ A=A^*$.

\begin{theorem}
Let $A:M\rightarrow M$ be a bicomplex self-adjoint operator and
$\ket{\psi}\in M$ be an eigenvector of the equation
$A\ket{\psi}=\lambda \ket{\psi}$, with $\ket{\psi}\notin
\mathcal{NC}$. Then the eigenvalues of $A$ are in the set of
hyperbolic numbers.
\end{theorem}

\noindent \emph{Proof.} If $A$ is a bicomplex self-adjoint
operator $A=A^*$ on $M$ and $A\ket{\psi}=\lambda \ket{\psi}$ with
$\lambda\in \mathbb{T}$ then
\begin{equation}
\bra{\psi}A\ket{\psi}=\lambda \braket{\psi}{\psi},
\label{paplp}
\end{equation}
where $\braket{\psi}{\psi}\in \mathbb{D}^+$. Moreover, we have
$$
\bra{\psi}A\ket{\psi}^{\dag_3}=\bra{\psi}A^*\ket{\psi}=\bra{\psi}A\ket{\psi}.
$$
This implies that $\bra{\psi}A\ket{\psi}\in \mathbb{D}$. Since
$\braket{\psi}{\psi}\notin \mathcal{NC}\Leftrightarrow\ket{\psi}\notin \mathcal{NC}$, we can divide each side
of equation (\ref{paplp}) by $\braket{\psi}{\psi}$. Therefore,
$\lambda$ can only be in $\mathbb{D}$. $\Box$

\smallskip
\noindent \textbf{Remark}. The requirement that the eigenvector
$\ket{\psi}$ is not in the null-cone means that $\ket{\psi}=\eo
\ket{\psi_{\eo}}+\et \ket{\psi_{\et}}$ with $\ket{\psi_{\eo}}\neq
\ket{0}$ \mbox{and} $\ket{\psi_{\et}}\neq \ket{0}$.

%\begin{corollary}
%Let $A:M\rightarrow M$ be a bicomplex self-adjoint operator and
%$\ket{\psi}\in V$ be an eigenvector of the equation
%$A\ket{\psi}=\lambda \ket{\psi}$, with $\ket{\psi}\neq \ket{0}$.
%Then the eigenvalues of $A$ are in the real numbers.
%\end{corollary}
%\emph{Proof}. This is a direct consequence of theorem \ref{AinV}.
%$\Box$

%Now, by a similar proof as the usual one we can show that two
%eigenvectors in $M$ of a bicomplex Hermitian operator associated
%with two different eigenvalues in $\mathbb{D}$ are orthogonal. In
%other words, if $A\ket{\psi}=\lambda \ket{\psi}$ and
%$A\ket{\phi}=\mu \ket{\phi}$ with $\lambda-\mu\notin
%\mathcal{NC}$, then $\braket{\phi}{\psi}=0$.

\end{document}